\documentclass[a4paper]{article}

\usepackage{INTERSPEECH2022}

\usepackage{multirow}
\usepackage{subfigure}
\usepackage{CJKutf8}
\usepackage{hyperref}

\title{WeNet 2.0: More Productive End-to-End Speech Recognition Toolkit}
\name{Binbin Zhang$^{1,3}$,  Di Wu$^{1,3}$, Zhendong Peng$^{1,3}$, Xingchen Song$^{1,3}$,
Zhuoyuan Yao$^{2}$, Hang Lv$^{2,3}$,  Lei Xie$^{2,*}$, Chao Yang$^{1,3}$, Fuping Pan$^1$, Jianwei Niu$^1$}

\address{
  $^1$Horizon Robotics, Beijing, China \\
  $^2$Audio, Speech and Language Processing Group (ASLP@NPU),
School of Computer Science, Northwestern Polytechnical University, Xi’an, China \\
  $^3$ WeNet Open Source Community \\
  \thanks{* Lei Xie is corresponding author.}
  }
\email{binbin.zhang@horizon.ai, hanglv@nwpu-aslp.org, lxie@nwpu.edu.cn}

\begin{document}

\maketitle

\begin{abstract}
Recently, we made available WeNet~\cite{2021WeNet}, a production-oriented end-to-end speech recognition toolkit, which introduces a unified two-pass (U2) framework and a built-in runtime to address the streaming and non-streaming decoding modes in a single model. To further improve ASR performance and facilitate various production requirements, in this paper, we present WeNet 2.0 with four important updates.
(1) We propose U2++, a unified two-pass framework with bidirectional attention decoders, which includes the future contextual information by a right-to-left attention decoder to improve the representative ability of the shared encoder and the performance during the rescoring stage.
(2) We introduce an n-gram based language model and a WFST-based decoder into WeNet 2.0, promoting the use of rich text data in production scenarios.
(3) We design a unified contextual biasing framework, which leverages user-specific context (e.g., contact lists) to provide rapid adaptation ability for production and improves ASR accuracy in both with-LM and without-LM scenarios.
(4) We design a unified IO to support large-scale data for effective model training.
In summary, the brand-new WeNet 2.0 achieves up to 10\% relative recognition performance improvement over the original WeNet on various corpora and makes available several important production-oriented features.

\end{abstract}


\section{Introduction}

End-to-end (E2E) methods, such as connectionist temporal classification (CTC)~\cite{graves2006connectionist}, recurrent neural network transducer (RNN-T)~\cite{graves2012sequence,wang2021cascade,wang2019exploring}, and attention based encoder-decoder (AED)~\cite{chorowski2014end,chan2015listen,luo2021simplified,miao2020transformer}, have drawn immense interest in the last few years in automatic speech recognition (ASR). Recent works~\cite{prabhavalkar2017comparison,sainath2019two,kim2017joint,sainath2020streaming,hu2020deliberation} show the E2E systems not only extremely simplify the speech recognition pipeline, but also surpass the conventional hybrid ASR systems in accuracy. 

Considering the advantages~\cite{li2021recent} of E2E models, deploying the emerging E2E ASR framework into real-world productions with stable
and highly efficient characteristics becomes necessary. However, almost all of the well-known E2E speech recognition toolkits, such as ESPnet~\cite{watanabe2018espnet} and SpeechBrain~\cite{ravanelli2021speechbrain}, are research-oriented rather than production-oriented.

In \cite{2021WeNet}, we present a production first and production ready E2E speech recognition toolkit, WeNet\footnote{\url{https://github.com/wenet-e2e/wenet}}, which focuses on addressing the production problems of transformer~\cite{vaswani2017attention} and conformer~\cite{gulati2020conformer} based E2E models.
Specifically, WeNet adopts a joint CTC/AED structure as the basic model structure. And then, a unified two-pass (U2) framework is proposed to solve the streaming problem, where a dynamic chunk masking strategy is applied in the training procedure to unify the streaming and non-streaming modes in a unified neural model. A built-in runtime is provided in which developers can run x86 and android E2E system for speech recognition out of the box.
WeNet dramatically reduces the workload of deploying an E2E model in real applications, so it is widely adopted by researchers and developers.

In this paper, we present \textbf{WeNet 2.0}, which introduces the recent developments and solutions updated in WeNet for production-oriented speech recognition. The key updates of WeNet 2.0 are as follows.

\textbf{(1) U2++}: we upgrade the original U2 framework to U2++, which simultaneously utilizes the left-to-right and right-to-left bidirectional contextual information of the labeled sequences to learn richer context information during the training stage, and combines the forward and backward predictions to achieve more accurate results during the decoding stage. The experiments show that U2++ achieves up to 10\% relative reduction over the original U2 framework in error rate.

\textbf{(2) Production language model solution}: we support an optional n-gram LM, which is composed with the E2E modeling unit in a weighted finite state transducer (WFST)~\cite{mohri1997finite} based decoding graph, during the streaming CTC decoding stage. The n-gram LM can be trained rapidly on rich production-accumulated text data. Experiments show that the n-gram language model can provide up to 8\% relative performance improvement.

\textbf{(3) Contextual biasing}: we design a unified contextual biasing framework which provides the opportunity to leverage the user-specific contextual information with or without LM during the streaming decoding stage. Utilizing user-specific contextual information (e.g., contact lists, particular dialog state, conversation topic, location, etc) plays a great role in both improving the ASR accuracy and providing rapid adaptation ability. Experiments show that our contextual biasing solution can bring clear dramatic improvement for both with-LM and without-LM cases.

\textbf{(4) Unified IO (UIO)}: we design a unified IO system to support the large-scale dataset for effective model training. The UIO system provides a unified interface for different storage media (i.e., local disk or cloud) and datasets for different scale (i.e., small or large datasets). For small datasets, UIO keeps the sample-level random access ability. While, for large datasets, UIO aggregates the data samples to shards (refer to  Figure \ref{fig:uio} for more details) and provides the shard-level random access ability. Thanks to UIO, WeNet 2.0 can elastically support training with a few hours to millions of hours of data.

In summary, our brand-new WeNet 2.0 makes available several important production-oriented features, which achieves substantial ASR improvement over the original WeNet and makes itself a more productive E2E speech recognition toolkit. 

\vspace{-5pt}
\section{WeNet 2.0}

In this section, we will describe the production-oriented key updates: U2++ framework, production language model solution, contextual biasing, and unified IO in detail, respectively.

\vspace{-2pt}
\subsection{U2++}
\begin{figure}[ht]
  \centering
  \includegraphics[width=\linewidth]{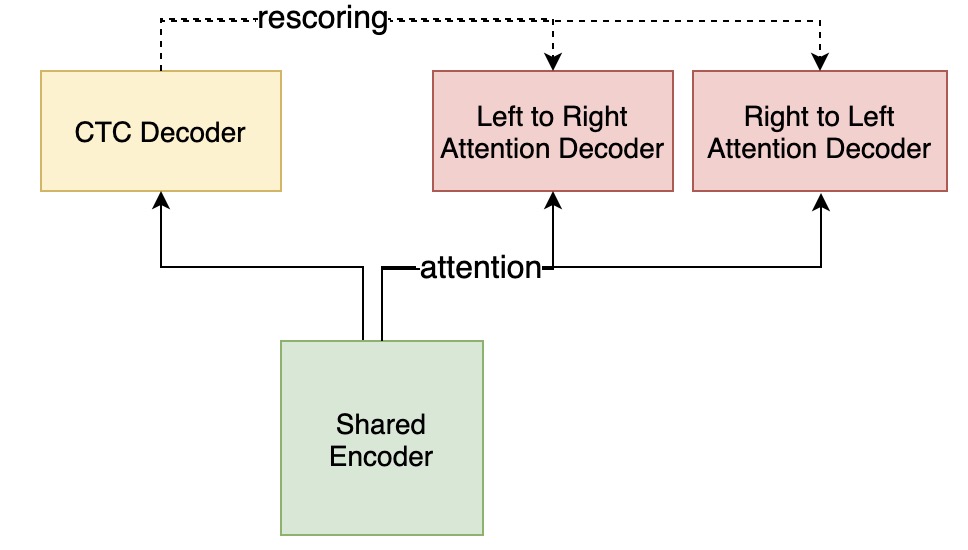}
  \caption{U2++ framework. The dotted line denotes the second pass rescoring during the decoding stage.}
  \label{fig:ctc_attention_joint}. 
  \vspace{-2em}
\end{figure}

U2++, a unified two-pass joint CTC/AED framework with bidirectional attention decoders, provides an ideal solution to unify streaming and non-streaming modes. As shown in Figure~\ref{fig:ctc_attention_joint}, U2++ consists of four parts.
1) A \textit{Shared Encoder} that models the information of acoustic features. The \textit{Shared Encoder} consists of multiple Transformer~\cite{vaswani2017attention} or Conformer~\cite{gulati2020conformer} layers which only takes limited right contexts into account to keep a balanced latency.
2) A \textit{CTC Decoder} that models the frame-level alignment information between acoustic features and token units. The \textit{CTC Decoder} consists of a linear layer, which transforms the \textit{Shared Encoder} output to the CTC activation.
3) A \textit{Left-to-Right Attention Decoder (L2R)} that models the ordered token sequence from left to right to represent the past contextual information. 
4) A \textit{Right-to-Left Attention Decoder (R2L)} that models the reversed token sequence from right to left to represent the future contextual information.
The \textit{L2R and R2L Attention Decoders} consist of multiple Transformer decoder layers.
During the decoding stage, the \textit{CTC Decoder} runs in the streaming mode in the first pass and the \textit{L2R and R2L Attention Decoders} do rescore in the non-streaming mode to improve the performance in the second pass.

Compared with U2~\cite{2021WeNet}, we add an extra right-to-left attention decoder to enhance the context modeling ability of our model, so that the context information comes not only from the past (left-to-right decoder) but also from the future (right-to-left decoder). This improves the representative ability of the shared encoder, the generalization ability of the whole system, and the performance during the rescoring stage.


The combined CTC and AED loss is used to train U2++:
\begin{equation}
\setlength\abovedisplayskip{2pt}
\setlength\belowdisplayskip{2pt}
  \label{eq:combined_loss}
    \mathbf{L}_{\text {combined}}\left(\mathbf{x}, \mathbf{y}\right)=\lambda \mathbf{L}_{\text{CTC}}\left(\mathbf{x}, \mathbf{y}\right)+(1-\lambda) (\mathbf{L}_{\text{AED }}\left(\mathbf{x}, \mathbf{y}\right)),
\end{equation}
where $\mathbf{x}$ denotes the acoustic features, and $\mathbf{y}$ denotes the corresponding labels, $\lambda$ is a hyper parameter which balances the importance of the CTC loss $\mathbf{L}_{\text{CTC}}\left(\mathbf{x}, \mathbf{y}\right)$ and AED loss $\mathbf{L}_{\text {AED}}\left(\mathbf{x}, \mathbf{y}\right)$.

To incorporate an extra R2L attention decoder into our model, we introduce a hyper parameter $\alpha$ inside $\mathbf{L}_{\text {AED}}\left(\mathbf{x}, \mathbf{y}\right)$ to adjust the contributions of two unidirectional decoders:
\begin{equation}
\setlength\abovedisplayskip{2pt}
\setlength\belowdisplayskip{2pt}
  \label{eq:AED_loss}
    \mathbf{L}_{\text {AED}}\left(\mathbf{x}, \mathbf{y}\right)=(1 - \alpha) \mathbf{L}_{\text{L2R}}\left(\mathbf{x}, \mathbf{y}\right) + \alpha (\mathbf{L}_{\text{R2L}}\left(\mathbf{x}, \mathbf{y}\right) ).
\end{equation}

Similar to U2, the dynamic chunk masking strategy is used to unify the streaming and non-streaming modes. During the training stage, we firstly sample a random chunk size $C$ from a uniform distribution, $C \sim U(1, max \, batch \, length \, T)$. And then, the input is split into several chunks with the chosen chunk size. At last, the current chunk does bidirectional chunk-level attention to itself and previous/following chunks in L2R/R2L attention decoder respectively in training. During the decoding stage, the n-best results achieved from the first-pass CTC decoder are rescored by the L2R and R2L attention decoder with its corresponding acoustic information generated by the \textit{Shared Encoder}. The final results are obtained by fusing the scores of the two attention decoders and the CTC decoder. 

Empirically, a larger chunk size results in better results with higher latency. Thanks to the dynamic strategy, U2++ learns to predict with arbitrary chunk size so that balancing the accuracy and latency is simplified by tuning the chunk size in decoding.




\vspace{-2pt}
\subsection{Language Model}

\begin{figure*}[ht]
  \centering
  \includegraphics[width=\linewidth]{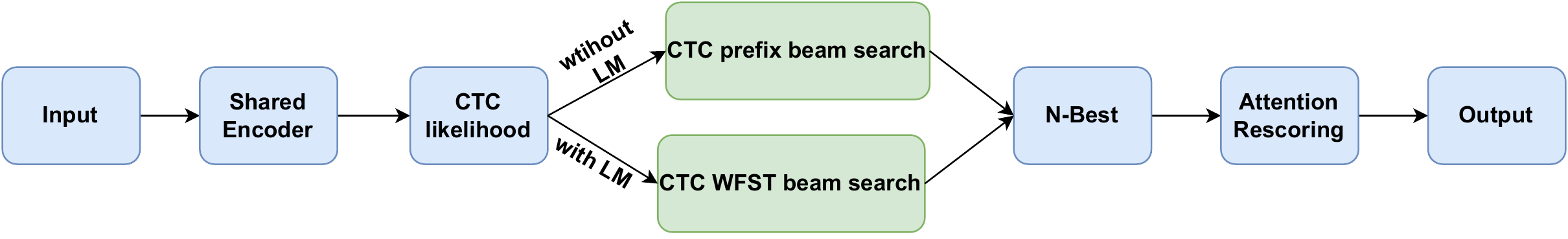}
  \vspace{-2em}
  \caption{Language model solution in WeNet 2.0}
  \label{fig:lm}
  \vspace{-2.0em}
\end{figure*}

To promote the use of rich text data in production scenarios, we provide a unified LM integration framework in WeNet 2.0, which is shown in Figure~\ref{fig:lm}.

As a unified LM and LM-free system, CTC is employed to generate the first-pass n-best results.
When LM is not provided, \textit{CTC prefix beam search} is applied to get n-best candidates.
When LM is supplied, WeNet 2.0 compiles the n-gram LM ($G$), lexicon ($L$), and end-to-end modeling CTC topology ($T$) into a WFST-based decoding graph ($TLG$)
\begin{equation}
    \setlength\abovedisplayskip{2pt}
    \setlength\belowdisplayskip{2pt}
    TLG = T \circ \min(det(L \circ G)),
\end{equation}
and then \textit{CTC WFST beam search} is applied to get n-best candidates. Finally, the n-best candidates are rescored by the \textit{Attenion Rescoring} module to find the best candidate.

We reuse the algorithm and code in Kaldi\cite{povey2011kaldi} for decoding, which is denoted as \textit{CTC WFST beam search} in the WeNet 2.0 framework. To speedup decoding, blank frame skipping~\cite{chen2016phone} technique is adopted.

\vspace{-2pt}
\subsection{Contextual Biasing}

Utilizing user-specific contextual information (e.g., contact lists, driver's navigation) plays a vital role in speech production, which always improves the accuracy significantly and provides the rapid adaptation ability. Contextual biasing technique is investigated in \cite{aleksic2015bringing, he2019streaming, jain2020contextual} for both traditional and E2E systems. 

\begin{figure}[ht]
\centering
    \subfigure[Char level context graph without LM]{
        \includegraphics[width=1.0\linewidth]{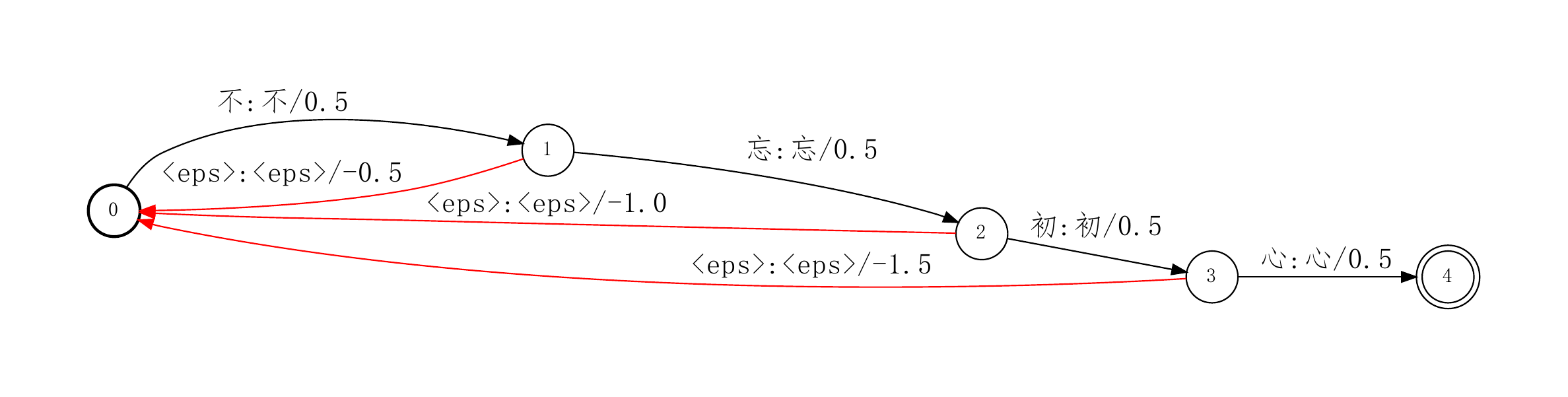}
    }
    \quad
    \subfigure[Word level context graph with LM]{
        \includegraphics[width=0.7\linewidth]{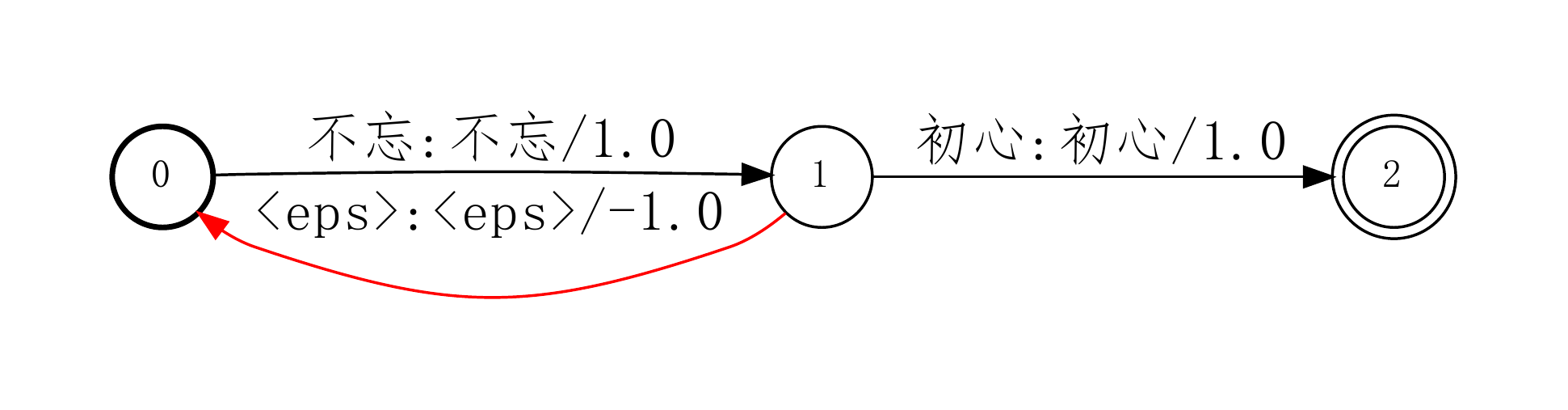}
    }
\vspace{-2pt}
\caption{Context graph for Chinese phrase "不忘初心". The failure arcs are highlighted in red lines.}
\label{fig:context}
\end{figure}

In WeNet 2.0, for leveraging the user-specific contextual information in both LM and LM-free cases during the streaming decoding stage, inspired by \cite{he2019streaming}, we construct a contextual WFST graph on the fly when a set of biasing phrases are known in advance. 
Firstly, the biasing phrases are split into biasing units according to the E2E modeling units in the LM-free case or the vocabulary words in the with-LM case.
And then, the contextual WFST graph is built on the fly as follows: (1) Each biasing unit with a boosted score is placed on a corresponding arc sequentially to generate an acceptable chain. (2) For each intermediate state of the acceptable chain, a special failure arc with a negative accumulated boosted score is added. The failure arcs are used to remove the boosted scores when only partial biasing units are matched rather than the entire phrase. In Figure~\ref{fig:context}, a char (E2E modeling unit) level context graph in the LM-free case and a word-level context graph in the with-LM case are shown, respectively. Finally, during the streaming decoding stage, a boosted score is added immediately when the beam search result matches with a biasing unit through the contextual WFST graph, as shown in
\begin{equation}
\setlength\abovedisplayskip{2pt}
\setlength\belowdisplayskip{2pt}
\label{equ:context}
\textbf{y}^{*} = \mathop{\arg\max}_{\textbf{y}} logP(\mathbf{y}|\mathbf{x}) + \lambda logP_{C}(\mathbf{y}),
\end{equation}
where $P_{C}(\mathbf{y})$ is the biasing score, $\lambda$ is a tunable hyper-parameter to control how much the contextual LM influences the overall model score. Especially, when some biasing phrases share the same prefix, we do a greedy match to simplify the implementation.


\vspace{-10pt}
\begin{CJK*}{UTF8}{gbsn}
\end{CJK*}
\vspace{-10pt}

\subsection{UIO}
Typically, the production-scale speech dataset is over tens of thousands of hours labeled speech, which always comprises massive small size files. The massive small files will cause the following problems:

\vspace{-3pt}
\begin{itemize}
\item \textbf{Out of memory (OOM)}: we have to keep the index information for all of the small files, which is very memory-consuming for large datasets.
\vspace{-3pt}
\item \textbf{Lower training speed}: reading data becomes the bottleneck of the training due to the intolerable time-consuming during random access to massive small files.
\end{itemize}
\vspace{-3pt}

\begin{figure}[h]
  \centering 
  \includegraphics[width=\linewidth]{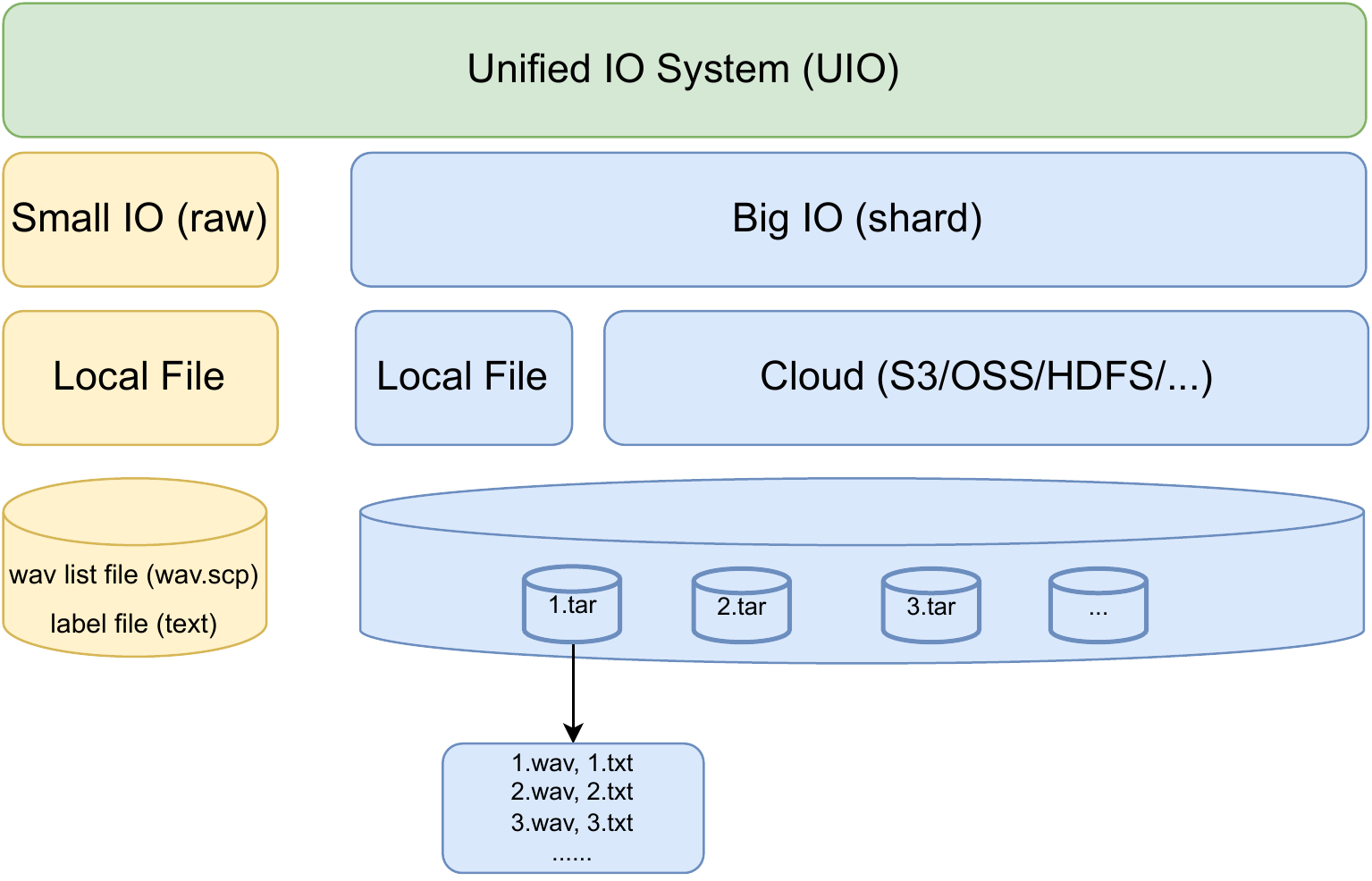}
  \caption{Unified IO in WeNet 2.0}
  \label{fig:uio}
  \vspace{-1.0em}
\end{figure}

To address the aforementioned problems for large-scale production datasets and keep the high efficiency for small datasets at the same time, we design a unified IO system, which provides a unified interface for different storage (i.e., local disk or cloud) and different scale datasets (i.e., small or large datasets). The whole framework is shown in Figure~\ref{fig:uio}.

For large datasets, inspired by TFReord (Tensorflow)~\cite{abadi2016tensorflow} and AIStore~\cite{aizman2019high}, we pack each set of samples (e.g., 1000 samples) with their associated metadata information into a corresponding bigger shard, which is done by GNU tar, an open-source, open-format, ubiquitous and widely-supported archival tool. Tar files significantly save memory and overcome the OOM problem effectively. During the training stage, on-the-fly decompression is performed in the memory, and the data in the same compressed shard is read sequentially, which solves the time-consuming random data access problem and speeds up the training process. At the same time, different shards can be read randomly to ensure the global randomness of data. For small datasets, we can also load the training samples directly.

Especially, when using shards for big datasets, we support loading the shards from both the local disk and the distributed storage (e.g., S3, OSS, HDFS, etc). Similar to TFRecord, chain operations for processing data on-the-fly is designed, so that the unified IO is extensible and easy to debug\footnote{\url{https://wenet.org.cn/wenet/UIO.html}}.

\section{Experiments}

In this section we describe our experimental setup, test sets, and analyze the experimental results. Most of the experimental setups are available in the WeNet recipes.

The experiments are carried on all of or some of the following corpora: AISHELL-1~\cite{bu2017aishell}, AISHELL-2~\cite{du2018aishell}, LibriSpeech~\cite{panayotov2015librispeech},  GigaSpeech~\cite{chen2021gigaspeech}, and recent released WenetSpeech~\cite{zhang2021wenetspeech}. The five corpora include different languages (English and Mandarin), recording environments (clean and noisy), and sizes (100 - 10000 hours).

\vspace{-3pt}
\subsection{U2++}
\label{sec:u2++exp}
\vspace{-1pt}

\begin{table*}[htbp]
\centering
\caption{U2++/U2 comparison on various open-source ASR corpora. * denotes the model is trained with dynamic mask, and it could be decoded with different chunk size (full/16 in the table). † denotes the model is trained with full attention.}
\vspace{-7pt}
\setlength{\tabcolsep}{3.5mm}
\label{tab:u2pp}
\begin{tabular}{@{}lllllccccc@{}}
\toprule
Dataset       & Language & Hours & Unit & Metric & Test Sets                   & U2                  & U2++                 \\ \midrule
AISHELL-1 *   & Mandarin & 170   & Char  & CER    & test(full)/test(16)         & 4.97/5.45           & \textbf{4.63/5.05}   \\
AISHELL-2 *   & Mandarin & 1000  & Char  & CER    & test ios(full)/test ios(16) & 6.08/6.46           & \textbf{5.39/5.78}   \\
LibriSpeech † & English  & 960   & BPE   & WER    & test clean/test other       & 2.85/7.24           & \textbf{2.66/6.53}   \\
WenetSpeech † & Mandarin & 10000 & Char  & MER    & Test Net/Test Meeting       & 9.70/\textbf{15.59} & \textbf{9.25}/16.18  \\
GigaSpeech †  & English  & 10000 & BPE   & WER    & dev/test                    & 11.30/11.20         & \textbf{10.70/10.60}         \\ \bottomrule
\end{tabular}
\vspace{-2em}
\end{table*}

To evaluate the effectiveness of our U2++ model, we conduct experiments on all of the 5 ASR corpora listed above. For most experiments, 80-dimensional log Mel-filter banks (FBANK) with a 25ms window and a 10ms shift are used as acoustic features.
SpecAugment~\cite{park2019specaugment} is applied on-the-fly for data augmentation.
Two convolution subsampling layers with kernel size 3*3 and stride 2 are used in the front of the encoder.
We use 12 conformer layers for the encoder.
To keep comparable parameters for U2/U2++, we use 6 transformer decoder layers for U2, 3 left-to-right and 3 righ-to-left decoder layers for U2++.
Moreover, we obtain our final model by model averaging.
In AISHELL-1 and AISHELL-2, the attention layer uses attention dimension 256, feed-forward 2048, and 4-head attention.
In LibriSpeech, GigaSpeech and WenetSpeech, the attention layer uses attention dimension 512, feed-forward 2048, and 8-head attention. 
The kernel size of the convolution module is 8/8/31/31/15 for the five corpora, respectively.
Accumulating gradient is used to stabilize training.

In Table~\ref{tab:u2pp}, we report the character error rate (CER), word error rate (WER) or mixed error rate (MER) for each corpus. It shows that U2++ outperforms U2 on most of the corpora, and achieves up to 10\% relative improvement over U2. From the results, we approve that U2++ shows superior performance in various types and sizes of ASR corpora consistently.

\vspace{-3pt}
\subsection{N-gram Language Model}
\vspace{-1pt}

The language model solution is evaluated on AISHELL-1, AISHELL-2, and LibriSpeech on the models listed in Section~\ref{sec:u2++exp}.
For AISHELL-1 and AISHELL-2, a tri-gram trained on its own training corpus is used.
For LibriSpeech, the official pre-trained four-gram large language model (fglarge) is used.
To speed up the decoding, when the probability of the ``blank'' symbol of a frame is greater than 0.98, the frame will be skipped.

\vspace{-1em}
\begin{table}[ht]
\caption{N-gram LM evaluation}
\vspace{-5pt}
\label{tab:lm}
\begin{tabular}{@{}lccc@{}}
\toprule
Dataset     & Test Sets               & without LM & with LM   \\ \midrule
AISHELL-1   & test                    & 4.63       & \textbf{4.40}      \\
AISHELL-2   & test\_ios               & 5.39       & \textbf{5.35}      \\
LibriSpeech & test\_clean/test\_other & 2.66/6.53  & \textbf{2.65}/\textbf{5.98} \\ \bottomrule
\end{tabular}
\end{table}
\vspace{-10pt}

Table \ref{tab:lm} exhibits the comparison results on the three corpora in with-LM and without-LM scenarios. The with-LM solution shows 5\% gain on AISHELL-1, comparable result on AISHELL-2, comparable result in test\_clean on LibriSpeech, and superior improvement (8.42\%) in test\_other on LibriSpeech.

In conclusion, our production LM solution shows consistent gain on the three corpora, which illustrates n-gram language model works well when integrated to an E2E speech recognition system.
Benefitting from this solution, the production system can enjoy the rich production-accumulated text data.

\vspace{-3pt}
\subsection{Contextual Biasing}

We evaluate contextual biasing in a contact scenario, and we design two test sets selected from AISHELL-2 test sets.

\vspace{-5pt}
\begin{itemize}
\item \textbf{test\_p}: positive test set with its relevant context. We select 107 utterances with person names, and all the names are attached as contextual biasing phrases in decoding.
\vspace{-5pt}
\item \textbf{test\_n}: negative test set without any contextual biasing phrases (i.e., names) in the positive testset. We randomly select 107 eligible utterances as the negative test set.
\end{itemize}
\vspace{-5pt}

We carry on the test on the AISHELL-1 U2++ model with and without LM.
We show that the strength of contextual biasing can be controlled by switching the boosted score. In our experiments, the boosted score is varied from 0 to 10, where 0 means context biasing is not applied.
As shown in Table~\ref{tab:context}, our solution reduces the error rate on the positive test set and a bigger boosted score usually gives better improvement. Furthermore, when a proper boosted score is set, we can improve the performance on the positive testset and avoid the performance degradation on negative test set at the same time.

\vspace{-8pt}
\begin{table}[ht]
\caption{Context Biasing evaluation}
\vspace{-6pt}
\centering
\label{tab:context}
\setlength{\tabcolsep}{2.5mm}
\begin{tabular}{@{}ccccccc@{}}
\toprule
\multirow{2}{*}{LM}  & \multirow{2}{*}{test set} & \multicolumn{5}{c}{boosted score}      \\ \cmidrule(l){3-7} 
                     &                           & 0     & 3     & 5     & 7     & 10     \\ \midrule
\multirow{2}{*}{NO}  & test\_p                   & 14.94 & 11.11 & 7.65  & \textbf{6.17}  & 6.54   \\
                     & test\_n                   & 7.45  & 7.45  & 7.45  & 7.57  & 8.06   \\ \midrule
\multirow{2}{*}{YES} & test\_p                   & 13.95 & 11.23 & 11.11 & 10.12 & \textbf{9.14}   \\
                     & test\_n                   & 7.20  & 7.20  & 7.20  & 7.20  & 7.33   \\ \bottomrule 
\end{tabular}
\end{table}
\vspace{-15pt}

\subsection{UIO}

We evaluate the performance of UIO on AISHELL-1~\cite{bu2017aishell} testset and WenetSpeech~\cite{zhang2021wenetspeech} testsets (i.e., test\_net/test\_meeting), including the accuracy, scalability and training speed.
Specifically, we use 8 GPUs on one machine for AISHELL-1, and 24 GPUs on three machines for WenetSpeech.

As shown in Table~\ref{tab:uio}, the UIO shows close accuracy in raw and shard modes on AISHELL-1, while the shard mode gets about 9.27\% speedup in terms of the training speed.
For WenetSpeech, since it is too slow to only use raw mode for training, the training speed is not shown. We adopt the result of ESPnet (marked with $\dagger$) as our baseline, which has the same configuration as our model.
The result of the model trained on WenetSpeech by shard mode is comparable to ESPnet, which further illustrates the effectiveness of UIO.

\vspace{-5pt}
\begin{table}[ht]
\caption{Unified IO evaluation}
\vspace{-5pt}
\label{tab:uio}
\begin{tabular}{llll}
\toprule
Dataset                      & metric           & raw                   & shard                             \\ \midrule
\multirow{2}{*}{AISHELL-1}   & speed(per epoch) & 593 sec               & \textbf{538 sec} \\
                             & WER              & 4.63                  & 4.67                              \\ \midrule
\multirow{2}{*}{WenetSpeech} & speed(per epoch) & NA                    & 75 min                            \\  
                             & WER              & 8.90/15.90($\dagger$) & 9.70/15.59                        \\ \bottomrule
\end{tabular}
\end{table}
\vspace{-15pt}



\section{Conclusion and Future Work}
In this paper, we proposed our more productive E2E speech recognition toolkit, WeNet 2.0, which introduces several important production-oriented features and achieves substantial ASR performance improvement.

We are working on WeNet 3.0, which mainly focuses on unsupervised self-learning, on-device model exploration and optimization, and other characteristics for production-level ASR.

\newpage

\scriptsize
\bibliographystyle{IEEEbib}
\newlength{\bibitemsep}\setlength{\bibitemsep}{.14\baselineskip plus .05\baselineskip minus .05\baselineskip}
\newlength{\bibparskip}\setlength{\bibparskip}{-1pt}
\let\oldthebibliography\thebibliography
\renewcommand\thebibliography[1]{%
  \oldthebibliography{#1}%
  \setlength{\parskip}{\bibitemsep}%
  \setlength{\itemsep}{\bibparskip}%
}
\bibliography{main}

\end{document}